\pdfoutput=1 % tells arXiv to build with pdfLaTeX
\documentclass[journal,twoside,web]{ieeecolor}
\usepackage{generic}
\usepackage{cite}
\usepackage{amsmath,amssymb,amsfonts}
\usepackage{algorithmic}
\usepackage{graphicx}
\usepackage{algorithm,algorithmic}
\usepackage{hyperref}
\hypersetup{hidelinks}
\usepackage{textcomp}
\def\BibTeX{{\rm B\kern-.05em{\sc i\kern-.025em b}\kern-.08em
    T\kern-.1667em\lower.7ex\hbox{E}\kern-.125emX}}
\begin{document}
\title{PhysiCal-PET: Physics-driven neural networks for the rapid position Calibration of PET Scanners}
\author{Mojahed Abushawish, Pablo Galve, \IEEEmembership{Member, IEEE}, Iván Colmenarejo Allavena, Carlos Galindo Morata, Sara Gaitan Dominguez and Jose Manuel Udias, \IEEEmembership{Senior Member, IEEE}
\thanks{This work was supported in part by the European Union through the
European Innovation Council's Pathfinder Open Programme under Grant 101099096
(RETIMAGER); in part by the Spanish Ministry of Science, Innovation and
Universities under the FPU Program under Grant FPU24/00918; and in part by
Grant JDC2023-051754-I funded by MCIU/AEI/10.13039/501100011033 and by the
European Social Fund Plus (ESF+).}
\thanks{This work did not involve human subjects or animals in its research.}
\thanks{MA, PG, ICA, CGM, SGD, and JMU are with the
Grupo de F\'{i}sica Nuclear, EMFTEL \& IPARCOS, Universidad Complutense de
Madrid, CEI Moncloa, 28040 Madrid, Spain, and also with the Health Research
Institute of the Hospital Cl\'{i}nico San Carlos (IdISSC), 28040 Madrid,
Spain.}
\thanks{PG is also with the Institute for Physical and Information
Technologies ``Leonardo Torres Quevedo'' (ITEFI), Spanish National Research
Council (CSIC), 28006 Madrid, Spain.}}

\maketitle

\begin{abstract}
Depth-of-interaction (DOI) information is essential for mitigating parallax error and preserving uniform spatial resolution in high-resolution and long-axial-field-of-view positron emission tomography (PET). Conventional DOI calibration relies on mechanically scanned collimated sources or detector-specific bench measurements, which become impractical as the number of detector modules grows. We present PhysiCal-PET, a physics-driven neural-network framework for fast, label-free DOI calibration of fully assembled PET scanners. Coincidence events are acquired with an uncollimated positron-emitting point source placed at a small number of known positions, and a network mapping DOI-sensitive detector signals to interaction depth is trained by minimizing the distance between the known source position and the line of response defined by the two predicted interaction positions, so that supervision is provided by the annihilation geometry alone, without measured DOI labels. In realistic simulations, the DOI mean absolute error decreased in parallel with the geometric loss and converged to 0.98~mm, confirming that the loss is a faithful surrogate for the unobservable DOI error. 
Applied to a preclinical scanner with 144 pixelated dual-ended-readout detector modules, the method calibrated the entire system from 11 manually placed source positions in 55~min of acquisition time, reducing the trimmed source-to-LOR distance by 24\%, improving the radial point-source FWHM from 4.99 to 2.90~mm (42\%), and raising the Derenzo peak-to-valley ratio of the 1.7~mm rods from 1.28 to 1.88 while separating the 1.2~mm rods, which remain merged in the uncalibrated reconstruction. Requiring no collimation, motion stages, or reference measurements, and scaling favorably with system sensitivity, PhysiCal-PET provides a practical route toward autonomous, routine DOI calibration of fully assembled PET scanners with a large number of detectors.
\end{abstract}

\begin{IEEEkeywords}
PET, depth of interaction (DOI), detector calibration, physics-driven deep learning, self-supervised learning.
\end{IEEEkeywords}

\section{Introduction}
\label{sec:introduction}
Positron emission tomography (PET) is a non-invasive molecular imaging modality that reconstructs the \emph{in vivo} distribution of a radiotracer from the pairs of 511~keV photons emitted in positron--electron annihilation~\cite{phelps2000positron}. PET detectors employ thick scintillation crystals to increase the detection efficiency of these photons. The absence of information about the depth at which each photon interacts introduces parallax error that degrades the spatial resolution~\cite{mohammadi2019minimization}. This effect is most severe in geometries where oblique lines of response (LORs) are common: small-diameter preclinical and organ-dedicated scanners~\cite{LEVIN2007125,lai2021h2rspet,Sanaat2024}, and total-body PET (TB-PET) systems with a long axial field of view~\cite{schmall2016parallax,vandenberghe2020state}. Measuring the depth of interaction (DOI) of each event mitigates parallax error and is therefore central to achieving a uniform, high spatial resolution across the entire field of view.

A broad range of detector concepts has been developed to encode DOI information, including phoswich crystals~\cite{seidel1999depth}, dual-ended readout~\cite{Kuang2020,Liu2025HighResolution}, staggered matrices of crystal layers~\cite{LIU2001182,GONZALEZ201682}, light-sharing and partial-light-sharing designs~\cite{Miyaoka}, and prismatoid light-guide arrays~\cite{labella2020high}; semi-monolithic and monolithic crystals additionally provide intrinsic DOI through the depth-dependent spread of scintillation light~\cite{gonzalezmontoro2021evolution}. In all cases the DOI is encoded in the detector signals rather than directly observed, so a calibration step is required to map each event's signal to an interaction depth.

This DOI calibration is regarded as an obstacle to deploying DOI-capable detectors in large systems. Conventional calibration provides reference DOI positions by irradiating the detector with collimated pencil- or fan-beams oriented laterally, so that depth is fixed by the beam geometry~\cite{borghi2015experimental,vandam2011practical}. The resulting signal--position pairs are then used to build a look-up table or to train a supervised model such as $k$-nearest-neighbor search, maximum-likelihood estimation, gradient tree boosting, or a neural network~\cite{Yang2009_doicalib,borghi2015experimental,mueller2018gradient,mueller2019novel,iborra2019ensemble,decuyper2021artificial}. Such procedures are accurate but slow, often requiring hours per detector with motorized scanning stages on dedicated benchtop setups before assembly, and lateral irradiation becomes infeasible once detectors are mounted in a closed ring. For systems with hundreds or thousands of modules, most acutely TB-PET, calibrating every module individually does not scale~\cite{vandenberghe2020state}.

To address scalability, recent work has focused on \emph{in-system} calibration, which calibrates detectors while they remain assembled in the scanner. Planar (transverse) in-system schemes based on virtual or electronic collimation can generate two-dimensional reference positions~\cite{gonzalezmontoro2020validation,dey2019vorrichtung}, but they recover DOI only parametrically from depth-dependent features of the light distribution, which does not fully exploit detectors with strong DOI encoding~\cite{vandam2011practical,mueller2022semi}. Methods that obtain genuine DOI labels in-system have therefore been proposed. Kuhl \textit{et al.} introduced angular fan-beam irradiation, in which oblique beams (and, in a dual-angle variant, their intersection) define the reference DOI of each event~\cite{kuhl2023angular}, and subsequently combined virtual collimation with angular irradiation to produce fully three-dimensional labels for gradient-tree-boosting models in an assembled proof-of-concept scanner~\cite{kuhl2025}. In a different direction, Yu \textit{et al.} exploited the intrinsic $^{176}$Lu radioactivity of LYSO to perform DOI self-calibration without any external source, matching a simulated intrinsic-radiation depth distribution to the measured dual-ended signal ratio~\cite{yu2026doi,alvasanchez2018understanding}; this removes the collimator but is restricted to lutetium-based scintillators and depends on the fidelity of the simulated detector response. Data-driven approaches have also been explored: Bae and Lee used unsupervised manifold learning (a Wasserstein-distance-based locally-linear embedding) to recover DOI from light distributions without labels~\cite{bae2026unsupervised,pedemonte2017machine}, although the method still requires perpendicular pencil-beam acquisitions and was demonstrated at the single-detector level.

Despite this progress, every existing approach still depends on at least one of the following: collimation hardware, mechanical scanning, the construction of an explicit geometric reference position (or pseudo-label) for each event, or a detector-specific simulated response. 

In this work we introduce PhysiCal-PET, a physics-driven, label-free framework for rapid DOI calibration of fully assembled PET scanners. Supervision is derived from the annihilation geometry alone: the DOI mapping is learned by enforcing consistency between each predicted line of response and a known source position, requiring no collimation, mechanical scanning, per-event depth labels, per-detector reference measurements, or detector-specific simulated response, and remaining agnostic to scintillator material, DOI-encoding mechanism, and readout technology. We validate the framework in realistic Monte Carlo simulations against ground truth and experimentally on a preclinical scanner with 144 dual-ended-readout modules, calibrating the entire system from 11 manually placed source positions in 55 min of acquisition time.

\section{Methods}
\label{sec:methods}

\subsection{Overview of the PhysiCal-PET Framework}
\label{subsec:method_overview}

PhysiCal-PET is a physics-driven, label-free framework for position calibration in fully assembled PET scanners. Coincidence events are acquired from an uncollimated positron-emitting point source placed at known positions within the field of view, and the calibration is supervised by the annihilation geometry alone: for each true coincidence, the two detected 511~keV photons define a LOR that must be consistent with the known source position. The framework therefore requires no collimated sources, detector-specific reference measurements, or ground-truth interaction-position labels.

In its general form, the framework estimates one or more unknown components of the gamma-ray interaction positions from the measured detector signals. For each coincidence event, the signals from the two interactions are passed to a neural network that predicts the unknown coordinates; these are combined with the known detector geometry to yield two three-dimensional interaction positions, whose connecting line defines the estimated LOR. The network is trained by minimizing the geometric inconsistency between this LOR and the known source position.

In this work, PhysiCal-PET is applied to DOI calibration in pixelated detectors: the transverse coordinates are fixed by the crystal segmentation, and the DOI is treated as the unknown coordinate to be estimated from DOI-sensitive signals.

The training strategy proposed here belongs to the broader family of physics-driven deep learning, in which the known physics of the imaging system, rather than a database of labeled examples, provides the supervisory signal. The general idea of deriving a training signal from physical laws was popularized by physics-informed neural networks, which embed governing equations directly in the loss~\cite{RAISSI2019686}. In medical imaging it is most often realized as physics-driven reconstruction, where networks enforce consistency with a known forward model; such unrolled, model-based methods are now among the most widely investigated deep-learning techniques for MRI~\cite{hammernik2018varnet,aggarwal2019modl} and have been extended to PET/MRI~\cite{reader2021petdl}. PhysiCal-PET applies this principle in the opposite direction: rather than using a calibrated system to reconstruct an unknown image, it uses a known image (a point source at a known location) to calibrate an unknown property of the system, namely the DOI response, with the annihilation geometry serving as the sole source of supervision.
In this respect PhysiCal-PET can be viewed as a generalization of PET normalization, in which a source of known distribution is acquired on the assembled scanner to characterize intrinsic detector properties such as crystal efficiencies and geometric factors~\cite{Badawi1999, Vicente2006}. Whereas component-based normalization estimates scalar, multiplicative sensitivity coefficients per crystal and per LOR from an extended uniform source, PhysiCal-PET estimates a physical interaction coordinate (the DOI) for each individual event through a learned nonlinear mapping. It thereby addresses a depth-dependent, non-stationary effect that multiplicative normalization cannot represent~\cite{Badawi1999}.

Throughout, ``label-free'' denotes the absence of per-event ground-truth interaction-depth labels; the known source positions act only as a geometric constraint shared by all events, not as a per-event reference.

\subsection{Neural Network Model}
\label{subsec:network_model}
The DOI estimator was implemented as a conditional neural network with shared weights that processes the two endpoints of each coincidence event independently. For the $n$th event, the network input is
\begin{equation}
    \left(
    \mathbf{x}_{n,1}, b_{n,1}, c_{n,1},
    \mathbf{x}_{n,2}, b_{n,2}, c_{n,2}
    \right),
\end{equation}
where, for endpoint $j \in \{1,2\}$, $\mathbf{x}_{n,j}$ is the detector signal vector, $b_{n,j}$ is the detector module index, and $c_{n,j}$ is the crystal index. Each crystal additionally has a fixed local transverse coordinate $\mathbf{p}_{n,j} = (x_{n,j}^{\mathrm{cryst}},y_{n,j}^{\mathrm{cryst}})$, determined by the known crystal segmentation. This coordinate is combined with the estimated DOI when reconstructing the global interaction position (Eq.~\ref{eq:local_to_global}).

Two input streams were formed. The detector signals were passed through a dense projection layer. In parallel, detector and crystal identity were encoded with learned embeddings: each physical crystal was assigned a global crystal index from its detector module and local crystal index. The detector embedding and global-crystal embedding were combined into a conditioning vector. The projected signal features were then processed by a series of feature-wise linear modulation (FiLM)~\cite{perez2018film} residual blocks~\cite{he2016deep}, in which the conditioning vector generated the scale and bias terms that modulate the intermediate features. This lets the DOI mapping depend on the specific detector module and crystal while keeping a single shared network across the scanner. The modulated features were then merged with the conditioning features and passed through a dense output head.

\begin{figure*}[!t]
\centerline{\includegraphics[width=\textwidth]{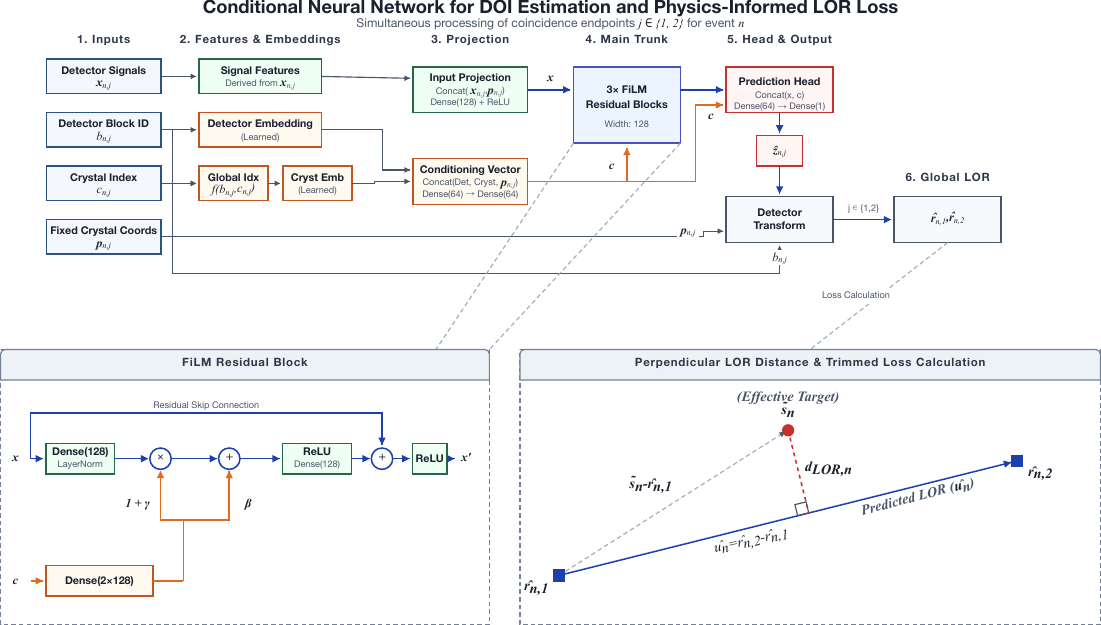}}
\caption{\textbf{Conditional neural network architecture for DOI estimation.} The two endpoints $j \in \{1,2\}$ of a coincidence event $n$ are processed in parallel by a shared-weight network. \textit{(Top)} Categorical inputs (detector module $b_{n,j}$ and crystal index $c_{n,j}$) are mapped to learned embeddings and projected by two $64$-unit dense layers into a conditioning vector; in parallel, the raw signal features are projected into a $128$-dimensional space. \textit{(Bottom left)} The trunk applies $3$ FiLM residual blocks, in which the conditioning vector is mapped by a dense layer to $256$ units, split into $128$-dimensional scale ($\gamma$) and bias ($\beta$) parameters. \textit{(Right)} A prediction head ($64$-unit dense layer to a $1$-unit output) yields the local DOI coordinate $\hat{z}_{n,j}$, which is combined with the fixed transverse coordinate $\mathbf{p}_{n,j}$ and transformed into global coordinates $\hat{\mathbf{r}}_{n,1}, \hat{\mathbf{r}}_{n,2}$. \textit{(Bottom right)} The model is trained end-to-end with the physics-driven trimmed LOR loss, which minimizes the perpendicular distance $d_{\mathrm{LOR},n}$ between the blurred source position $\tilde{\mathbf{s}}_n$ and the predicted LOR.}
\label{fig:diagram}
\end{figure*}

The network output for each endpoint was a scalar DOI estimate,
\begin{equation}
    \hat{z}_{n,j}
    =
    f_{\boldsymbol{\theta}}
    \left(
    \mathbf{x}_{n,j}, b_{n,j}, c_{n,j}
    \right),
    \qquad j \in \{1,2\},
\end{equation} 
where $f_{\boldsymbol{\theta}}$ denotes the network with trainable parameters $\boldsymbol{\theta}$. Combining $\hat{z}_{n,j}$ with the fixed transverse crystal coordinates and applying the known detector transformation gives the global interaction position. For detector module $b$ with origin $\mathbf{R}_b$ and local basis vectors $\mathbf{v}_{b,1}$, $\mathbf{v}_{b,2}$, $\mathbf{v}_{b,3}$,
\begin{equation}
    \hat{\mathbf{r}}_{n,j}
    =
    \mathbf{R}_{b_{n,j}}
    +
    \mathbf{v}_{b_{n,j},1}x_{n,j}^{\mathrm{cryst}}
    +
    \mathbf{v}_{b_{n,j},2}y_{n,j}^{\mathrm{cryst}}
    +
    \mathbf{v}_{b_{n,j},3}\hat{z}_{n,j}.
\label{eq:local_to_global}
\end{equation}
The resulting interaction positions $\hat{\mathbf{r}}_{n,1}$ and $\hat{\mathbf{r}}_{n,2}$ define the estimated LOR and are used to compute the distance-to-LOR loss. Fig.~\ref{fig:diagram} summarizes this architecture.

\subsection{Physics-Driven LOR Loss}
\label{subsec:loss_function}

The network was trained on the distance between the known source position and the LOR defined by the two predicted interaction positions. For an event from a source at $\mathbf{s}_n$ with predicted global positions $\hat{\mathbf{r}}_{n,1}$ and $\hat{\mathbf{r}}_{n,2}$, the predicted LOR direction is
\begin{equation}
    \hat{\mathbf{u}}_n
    =
    \hat{\mathbf{r}}_{n,2}
    -
    \hat{\mathbf{r}}_{n,1},
\end{equation}
and the perpendicular distance from the source to the LOR is
\begin{equation}
    d_{\mathrm{LOR},n}
    =
    \frac{
    \left\|
    \hat{\mathbf{u}}_n
    \times
    \left(
    \tilde{\mathbf{s}}_n
    -
    \hat{\mathbf{r}}_{n,1}
    \right)
    \right\|
    }{
    \left\|
    \hat{\mathbf{u}}_n
    \right\|
    }.
\label{eq:lor_distance}
\end{equation}

Here $\tilde{\mathbf{s}}_n$ is an effective source position obtained by perturbing the nominal coordinate,
\begin{equation}
    \tilde{\mathbf{s}}_n
    =
    \mathbf{s}_n
    +
    \boldsymbol{\eta}_n,
    \qquad
    \boldsymbol{\eta}_n
    \sim
    \mathcal{U}(-\Delta_s,\Delta_s),
\end{equation}
with blurring length $\Delta_s$ in each direction. This blurring accounts for the finite extent of the source, the uncertainty in its placement, and physical factors such as positron range or annihilation photon acollinearity, preventing the model from being optimized against a point-source location more precise than the acquisition geometry warrants.

To reduce the influence of non-ideal events including random coincidences, inter-crystal scatter, multiple-interaction events, mispositioned events, and events with incorrect crystal assignment, a trimmed loss~\cite{KARIMI2020101759} was used. Such events tend to produce large, uninformative source-to-LOR distances that would otherwise dominate the optimization. For each batch $\mathcal{B}$, the kept set was
\begin{equation}
    \mathcal{K}
    =
    \left\{
    n \in \mathcal{B}
    \;:\;
    d_{\mathrm{LOR},n} \leq d^{(K)}
    \right\},
    \quad
    K = \lfloor(1-\alpha)|\mathcal{B}|\rfloor,
\end{equation}
where $d^{(K)}$ is the $K$th smallest LOR distance in the batch and $\alpha \in [0,1)$ is the trimming factor; here $\alpha = 0.2$ excludes the largest 20\% of distances. The batch loss was
\begin{equation}
    \mathcal{L}_{\mathrm{batch}}
    =
    \frac{1}{|\mathcal{K}|}
    \sum_{n \in \mathcal{K}}
    d_{\mathrm{LOR},n}.
\label{eq:final_training_loss}
\end{equation}

The trimming factor $\alpha = 0.2$ was set to exceed the expected combined fraction of randoms, inter-crystal scatter, and mispositioned events. Because the trimming is applied per batch, the effective objective depends weakly on batch composition; with a batch size of 500 the trimmed quantile is estimated stably, and no sensitivity of the converged calibration to moderate variations of $\alpha$ was observed.

Minimizing this loss requires accurate source positions, a correct scanner geometry, sufficient statistics, and adequate angular diversity; in particular, oblique photon incidence increases the sensitivity of the loss to DOI errors, which motivated placing the calibration sources at a radial offset from the scanner center (Sections~\ref{subsec:simulation_setup} and~\ref{subsec:experimental_setup}).

\subsection{Scanner Geometry and Detector Configuration}
\label{subsec:scanner_geometry}

PhysiCal-PET was evaluated on a prototype preclinical cylindrical PET scanner with an inner ring diameter of 166~mm and an axial length of 152.6~mm, comprising 144 LYSO-based detector modules arranged in 6 rings of 24 blocks. Each block contained a $16 \times 16$ pixelated scintillator array of overall dimensions $22 \times 22 \times 15$~mm$^3$ (crystal pitch $\approx 1.375$~mm), with the 15~mm thickness defining the DOI direction.

Each block used an asymmetric dual-ended readout to encode DOI~\cite{Liu2025HighResolution}. One side was coupled to an array of silicon photomultipliers (SiPMs) of $3 \times 3$~mm$^2$ active area, multiplexed by a discretized positioning circuit (DPC)~\cite{park2022} into four position-encoding channels; the opposite side was coupled to an array of SiPMs of $6 \times 6$~mm$^2$ active area, summed in analog into a single channel. The blocks were oriented with the analog-summed side, hereafter referred to as the front side, facing the bore. Because the fraction of scintillation light reaching each side depends on interaction depth, the relative signal between the two sides provides a DOI-sensitive measurement that the network can learn.

\subsection{Simulation Study}
\label{subsec:simulation_setup}

A simulation study was performed to test whether the physics-driven objective could recover DOI under controlled conditions. The simulated geometry matched the physical scanner of Section~\ref{subsec:scanner_geometry}, including its 144 dual-ended readout blocks. Scanner-level emission and detector-interaction data were generated with UMC-PET, and optical light transport and SiPM signal formation were simulated with GOSS.

The purpose of this simulation study was solely to validate the PhysiCal-PET training strategy, i.e., to verify that an accurate DOI mapping can be learned from the distance-to-LOR loss alone by comparing against ground-truth interaction depths. It was not intended to replicate the experimental validation, and no quantitative agreement between simulated and measured results should be inferred.

UMC-PET (Ultra-fast Monte Carlo PET simulator)~\cite{galve2024umcpet} is a GPU-accelerated Monte Carlo simulation framework developed in CUDA-Fortran that models the full physics of PET emission, transport, and detection. By employing a high-resolution voxelized scheme, UMC-PET can accurately represent arbitrary scanner geometries, patient bodies, and complex detector structures. The simulator intrinsically models positron range, photon acollinearity, physical scatter, and attenuation within both the imaged object and the detector materials. In this study, UMC-PET was used to score the 3D position and deposited energy of each gamma interaction within the scintillator array, generating the input data required for the subsequent optical simulation.

To model the conversion of deposited energy into scintillation light and the subsequent signal collection, we employed GOSS (GPU-based Optical Simulator of Scintillator Crystals)~\cite{galve2024goss}. GOSS is a fast, CUDA-Fortran-based optical transport simulator focused on pixelated scintillator arrays with no restrictions regarding internal geometry or photosensor arrangement. It receives the gamma-interaction outputs from UMC-PET and tracks the generated optical photons via straight-line propagation from surface to surface. The optical transport across complex interfaces is modeled macroscopically using probabilistic parameters such as surface transparency, absorption, and roughness, which are empirically derived. The simulator additionally accounts for bulk properties like the material's photon attenuation and, the light yield as well as the SiPM photon detection efficiency. Accelerated by GPUs, GOSS enables efficient and realistic mapping of the light spread into the simulated dual-ended SiPM arrays.

The simulation placed an uncollimated point source at the same 11 field-of-view positions used in the experimental validation described in Section~\ref{subsec:experimental_setup}. For each position, $10^7$ positron decays were simulated; coincidences were retained only if both photons fell within the 511~keV photopeak window, yielding approximately $6.2 \times 10^6$ events. The network input consisted of the pre-multiplexing signals (the individual signal of each $3\times3$~mm$^2$ SiPM and the single summed signal of the front side), allowing PhysiCal-PET to be evaluated independently of the DPC readout electronics.

The dataset was split into training (99\%) and validation (1\%). Because the loss imposes an underdetermined geometric constraint rather than per-event DOI labels, conventional sample memorization is unlikely; the validation loss was therefore used to monitor optimization stability rather than generalization, which is assessed through the independent consistency checks of Sections~\ref{subsec:experimental_doi_distribution}--\ref{subsec:phantom_reconstruction}. The network was trained for $6 \times 10^{4}$ iterations with the distance-to-LOR loss (Section~\ref{subsec:loss_function}) using the Adam optimizer~\cite{kingma2014adam} with a learning rate of $10^{-3}$ and a batch size of 500. Because ground-truth coordinates were available, the DOI estimates were additionally evaluated with the DOI mean absolute error, used for assessment only and not during training.
To quantify the accuracy attainable when ground-truth interaction depths are available, a supervised reference model was additionally trained. This model used the identical network architecture, inputs, optimizer, learning rate, batch size, and iteration budget as the physics-driven model, and differed only in its objective: in place of the trimmed distance-to-LOR loss, it minimized the mean absolute error between the predicted DOI and the simulated ground-truth DOI. Because this model is optimized directly against the evaluation metric using perfect labels, its converged DOI MAE serves as a lower bound (floor) on the error achievable for the given detector signals and network capacity, against which the label-free result is compared.

\subsection{Experimental Calibration Acquisition}
\label{subsec:experimental_setup}

Experimental validation used the physical scanner of Section~\ref{subsec:scanner_geometry}. An uncollimated $^{22}$Na point source (nominal diameter 0.25~mm, activity $\approx 281~kBq$ at measurement) was placed at 11 field-of-view positions, each acquired for 5~min for a total of approximately 55~min. The positions sat at a radial offset of 50~mm from the scanner axis, increasing the fraction of oblique coincidences and thus the loss sensitivity to DOI errors (Section~\ref{subsec:loss_function}), with axial coordinates on three planes relative to the scanner center: four at $z=-32$~mm, four at $z=17$~mm, and three at $z=65$~mm.

Because the source was positioned manually, the nominal placement was subject to uncertainties on the order of several millimeters. To mitigate this, the geometric centroid of the source at each position was estimated by performing an initial image reconstruction with the DOI fixed at 5~mm. This depth was selected based on a preliminary analysis of a simulation of the scanner. Among the range of fixed-DOI values evaluated, a fixed depth of 5~mm provided the most accurate localization of the off-axis point sources. This value is also physically motivated: it is close to the mean
interaction depth ($\approx$5.9~mm) expected for 511~keV photons in
15~mm of LYSO, given a linear attenuation coefficient of
$\approx$0.87~cm$^{-1}$. Under this setting, the fixed-DOI reconstruction recovers the source coordinate with an accuracy of approximately 1~mm. The reconstructed centroids were subsequently used as the source positions $\mathbf{s}_n$ in the loss function defined in Eq.~\eqref{eq:lor_distance}.

Detector signals were then calibrated and events filtered. On each block, the summed four DPC channels and the analog-summed front-side signal were calibrated per crystal against the 511~keV photopeak. Events were retained only if both endpoints fell within a 400--600~keV window and the coincidence multiplicity was at most two. Events whose fixed-DOI LOR lay more than 9~mm from the reconstructed source position were also excluded, rejecting coincidences inconsistent with the nominal point-source geometry, such as randoms, scatter, or events with incorrect crystal assignment.

The input to the experimental neural network was constructed to expose both the energy-calibrated and uncalibrated detector responses to the model. For each endpoint, the signal vector $\mathbf{x}_{n,j}$ had seven components: the energy-calibrated DPC sum, the energy-calibrated front-side signal, the four raw DPC ADC values, and the raw front-side ADC value. Including both the calibrated and raw signals provided the network with energy-normalized features while preserving the underlying light-sharing information encoded in the individual DPC channels, which carries the DOI dependence.

The experimental network was trained using the distance-to-LOR loss and the Adam optimizer in two stages. In the first stage, the model was trained for $5 \times 10^{5}$ iterations with a source-position blurring parameter of $\Delta_s = 1$~mm and a learning rate of $10^{-3}$. This initial stage was designed to provide a robust coarse calibration while accounting for uncertainty in the nominal source positions. The trained network was then applied to the selected coincidence events to estimate DOI information, which was used to reconstruct the calibration source positions.

In the second stage, the reconstructed source positions from the first-stage DOI estimates were used to refine the calibration. The model was trained for a further $5 \times 10^{6}$ iterations with a reduced blurring parameter of $\Delta_s = 0.5$~mm and a learning rate of $10^{-4}$. The $\Delta_s$ value was chosen as a conservative estimate of the irreducible blurring imposed by the finite source size (0.25~mm diameter), the positron range in acrylic ($\approx 0.23$~mm~\cite{Chatziioannou1999MicroPET}), and photon acollinearity ($\approx 0.365$~mm~\cite{MOSES2011S236}), which together set a physical floor on the achievable source-to-LOR distance that no DOI estimate can reduce below. This second stage provided a finer optimization of the DOI mapping after reducing the effective uncertainty in the source-position estimates. The batch size used for training was 500.

Unlike the simulation study, the experimental data did not provide ground-truth DOI labels. Therefore, training convergence and calibration quality were assessed using indirect metrics: the trimmed distance-to-LOR loss, the mean distance-to-LOR, the predicted DOI distribution, the relationship between the predicted DOI and the dual-ended signal ratio, and the reconstructed spatial resolution at the calibration source positions.

\subsection{Signal-Ratio Analysis}
\label{subsec:signal_ratio_analysis}

As an experimental consistency check, the predicted DOI was compared with the ratio of the DPC-side signal to the total dual-ended signal~\cite{Liu2025HighResolution},
\begin{equation}
    R_{\mathrm{DPC}}
    =
    \frac{S_{\mathrm{DPC}}}{S_{\mathrm{DPC}} + S_{\mathrm{front}}},
\end{equation}
where $S_{\mathrm{DPC}}$ is the signal on the DPC-multiplexed side and $S_{\mathrm{front}}$ the analog-summed front-side signal. For selected crystals and blocks, the predicted DOI was plotted against $R_{\mathrm{DPC}}$. Because DOI~=~0~mm is defined at the front side, larger DOI should correspond to a greater relative DPC-side contribution; recovering this trend confirms that the learned mapping reflects the detector's light-sharing physics.

\subsection{Image Reconstruction}
\label{subsec:reconstruction_methods}

To assess the image-domain impact of the calibration, the 11 point-source measurements were reconstructed both with the DOI predicted by the trained network and with a fixed DOI of 5~mm (the non-calibrated baseline), using identical reconstruction settings otherwise. 

A standard OSEM~\cite{Hudson1994} reconstruction algorithm with image-based point spread function (PSF)~\cite{Arias-Valcayo2023} was implemented in-house. 

Spatial resolution was quantified from the reconstructed point-spread function. For each source, one-dimensional profiles along the radial, tangential, and axial directions were fitted with a Gaussian to obtain the full width at half maximum (FWHM), and the three FWHM values were compared across the 11 positions to quantify the resolution change from the DOI estimates.

A Derenzo phantom with rod diameters of 2.4, 2.0, 1.7, 1.2, 1.0 and 0.7 mm was placed 36 mm off the radial center and acquired for 30 min with an initial activity of 9.25 MBq. Events within an energy window of 400–600 keV were retained, yielding 338 million coincidences.

Rod separation was quantified from line profiles drawn along the rod rows of the 1.7 and 1.2~mm sections, in a direction close to the radial direction of the scanner: two rows in the 1.7~mm section (crossing 3 and 2 rods) and three in the 1.2~mm section (3, 3 and 2 rods), giving 3 and 5 inter-rod valleys respectively. Each profile was normalized to its maximum. For each valley, the peak-to-valley ratio (PVR) was defined as the mean amplitude of the two flanking peaks divided by the valley amplitude, and is reported per section as the mean $\pm$ standard deviation over its valleys. A valley was considered resolved when its amplitude did not exceed 73.5\% of the mean flanking peak amplitude (PVR $\geq$ 1.36), the contrast dip of the Rayleigh criterion~\cite{born2013principles}. Identical profiles were used for the DOI-based and fixed-DOI reconstructions, so the two are compared valley by valley.
\section{Results}
\label{sec:results}

\subsection{Simulation Training and DOI Accuracy}
\label{subsec:simulation_results}

The proposed physics-driven training strategy was first evaluated using simulated coincidence data, for which the ground-truth interaction depth was available. The simulation dataset contained approximately $6.2 \times 10^{6}$ preprocessed photopeak coincidence events, split into 99\% training and 1\% validation as described in Section~\ref{subsec:simulation_setup}.

\begin{figure}[t]
    \centering
    \includegraphics[width=\linewidth]{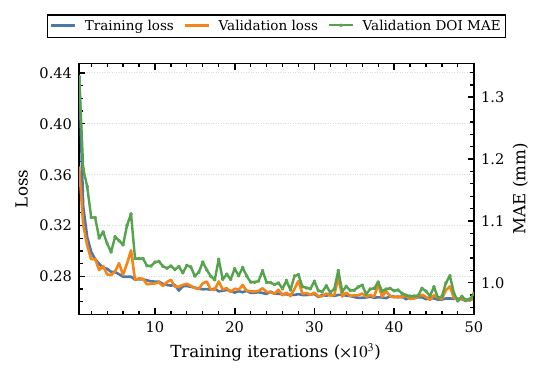}
    \caption{
    Training history of PhysiCal-PET in simulation. The training loss (blue line) and validation loss (orange line) are plotted against the left vertical axis, while the validation DOI mean absolute error (MAE) (green line) is plotted against the right vertical axis. All metrics are shown as a function of training iterations.
    }
    \label{fig:sim_training}
\end{figure}

Fig.~\ref{fig:sim_training} shows the training and validation curves. Both losses decreased consistently throughout optimization and converged to similar values, indicating stable training and the absence of overfitting. Crucially, the DOI mean absolute error (MAE), computed against the simulated ground truth, decreased in lockstep with the LOR-based loss even though it was never used as a training signal. This co-decrease is the central result of the simulation study: it demonstrates that minimizing the source-to-LOR distance provides effective supervision for DOI calibration in the absence of explicit DOI labels, and that the geometric loss is a surrogate for the DOI error.

After convergence, the physics-driven model reached a trimmed mean distance-to-LOR of 0.26~mm and a DOI MAE of 0.98~mm. To place this label-free accuracy in context, a supervised reference model with the identical architecture was trained directly against the ground-truth interaction depth (Section~\ref{subsec:simulation_setup}). This model, which is optimized against the quantity of interest using perfect labels, attained a DOI MAE of 0.96~mm and therefore represents the best accuracy achievable for the given detector signals and network capacity. The physics-driven model came within 0.02~mm (approximately 2$\%$) of this supervised floor despite never receiving a single DOI label during training. The near-equality of the two values indicates that the trimmed distance-to-LOR objective recovers almost all of the depth information encoded in the detector signals.

\begin{figure}[t]
    \centering
    \includegraphics[width=\linewidth]{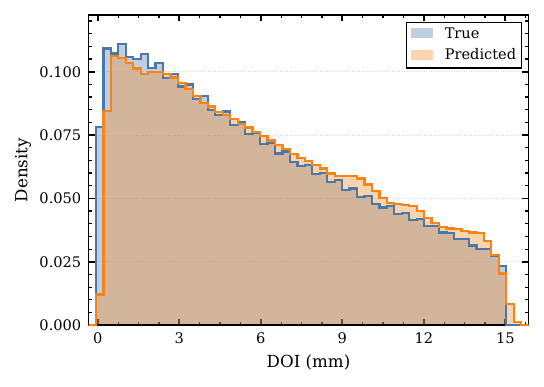}
    \caption{The simulated ground-truth DOI distribution (in blue) is compared with the DOI distribution predicted by PhysiCal-PET (in orange).}
    \label{fig:doi_distribution_sim}
\end{figure}

The distribution of the predicted DOI values was compared with the simulated ground-truth DOI distribution. As shown in Fig.~\ref{fig:doi_distribution_sim}, the predicted DOI distribution reproduced the main structure of the true interaction-depth distribution. This agreement confirms that the network learned a physically meaningful DOI mapping from the detector signals using only the distance-to-LOR constraint and the known source positions.

\subsection{Experimental Training Convergence}
\label{subsec:experimental_training}
Experimental validation was performed using coincidence data acquired from the preclinical PET scanner. The calibration dataset contained approximately $5.3 \times 10^{7}$ coincidence events acquired from 11 predefined source positions over a total duration of approximately 55~min.

\begin{figure}[t]
    \centering
    \includegraphics[width=\linewidth]{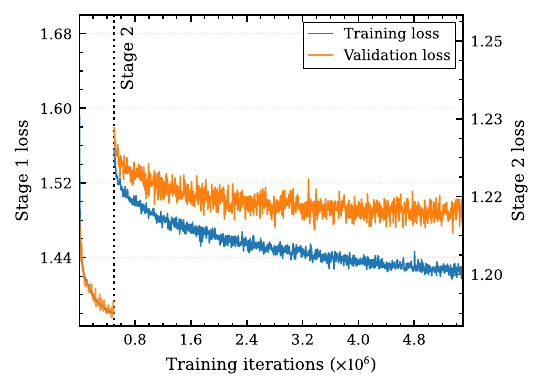}
    \caption{
    Experimental training convergence. The model was trained using coincidence events acquired from an uncollimated source at 11 positions.
    }
    \label{fig:experimental_training}
\end{figure}

Fig.~\ref{fig:experimental_training} shows the experimental training curve. The trimmed-mean LOR loss decreased steadily during optimization and converged to 1.20 mm. Note that the training loss is evaluated against the randomly blurred source positions ($\Delta_s = 0.5~mm$) and therefore slightly overestimates the geometric residual; all distance-to-LOR values reported below are computed against the unblurred reconstructed source positions. Since experimental ground-truth DOI labels were not available, performance was assessed using indirect validation metrics: the trimmed-mean distance-to-LOR, the predicted DOI distributions, the relationship between predicted DOI and detector signal ratio, and the reconstructed spatial resolution. The latter three are presented in Sections~\ref{subsec:experimental_doi_distribution}--\ref{subsec:point_source_reconstruction}.

To quantify the improvement provided by the learned calibration, the trimmed-mean distance-to-LOR was compared before and after training using the same trimming factor ($\alpha = 0.2$) as in the loss. With DOI fixed at 5~mm, the trimmed-mean distance-to-LOR was 1.49~mm. After training, this value decreased to 1.14~mm, corresponding to a 24\% reduction and indicating improved consistency between the predicted LORs and the known source positions. The untrimmed mean distance-to-LOR showed the same trend, decreasing from 2.23~mm with the fixed DOI to 1.92~mm after training, a 14\% reduction. The smaller relative improvement is expected, since the untrimmed mean remains dominated by the non-ideal coincidences excluded by the trimming.

\subsection{Experimental DOI distribution}
\label{subsec:experimental_doi_distribution}

\begin{figure}[t]
    \centering
    \includegraphics[width=\linewidth]{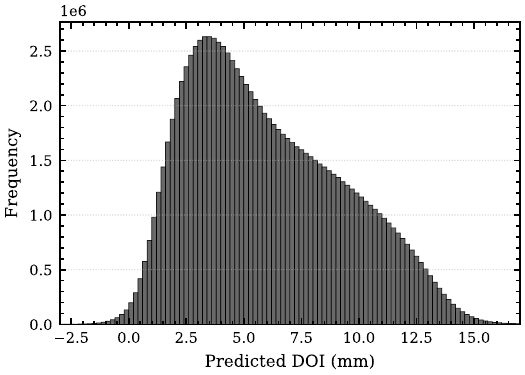}
    \caption{
    Experimental predicted DOI distribution obtained after training PhysiCal-PET on measured coincidence data. The predicted DOI values span the physical detector thickness and are consistent with the DOI behavior observed in simulation.
    }
    \label{fig:experimental_doi_distribution}
\end{figure}

The DOI distribution of the entire scanner predicted from the experimental data is shown in Fig.~\ref{fig:experimental_doi_distribution}. The distribution is physically plausible and is consistent with the expected attenuation of 511~keV photons within the 15~mm LYSO crystal thickness. Most predicted DOI values fall within the physical crystal depth, even though the network output was not explicitly constrained to this interval. This behavior indicates that the model learned a DOI mapping from the DOI-sensitive detector signals and the LOR-consistency objective.

\subsection{Relationship Between Predicted DOI and Dual-Ended Signal Ratio}
\label{subsec:ratio_doi_results}

\begin{figure}[t]
    \centering
    \includegraphics[width=\linewidth]{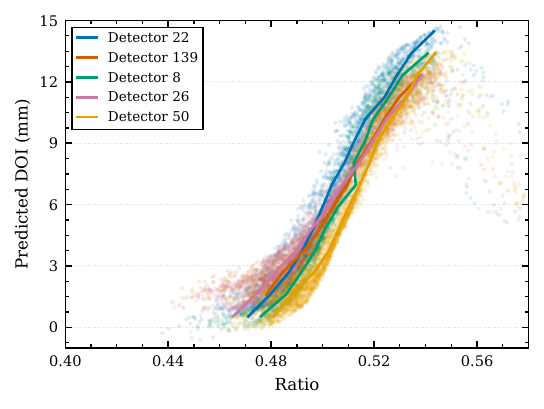}
\caption{
Relationship between predicted DOI and the dual-ended signal ratio $S_{\mathrm{DPC}}/(S_{\mathrm{DPC}}+S_{\mathrm{front}})$ for a selected crystal across five detector modules. Higher predicted DOI values correspond to increased relative signal contribution from the DPC side, consistent with the detector readout geometry. Solid lines indicate the median ratio within 1~mm predicted DOI bins.
}
    \label{fig:ratio_doi}
\end{figure}

To verify that the predicted DOI values were consistent with the physical response of the dual-ended detector readout, the predicted DOI was compared with the signal ratio defined in Section \ref{subsec:signal_ratio_analysis}.
Fig.~\ref{fig:ratio_doi} shows the predicted DOI as a function of this ratio for the same selected crystal across five different detector modules. A monotonic relationship was observed between the predicted DOI and the signal ratio. Higher DOI values corresponded to a larger relative signal contribution from the DPC side, which is consistent with the detector orientation used in this work, where DOI~=~0~mm was defined at the front side.

The figure also shows that, even for the same crystal index, the DOI--ratio relationship can vary slightly across detector modules. These detector-dependent variations highlight the need for a calibration method that can account for block-specific and crystal-specific response differences, rather than assuming a single global DOI--signal relationship for the full scanner.

This result provides an experimental consistency check for the learned calibration. Although the model was trained without DOI labels, the predicted DOI values followed the expected dependence on the dual-ended light-sharing signal.

\subsection{Point-Source Reconstruction With and Without DOI Calibration}
\label{subsec:point_source_reconstruction}

The effect of the learned DOI calibration on reconstructed spatial resolution was evaluated using the 11 point-source measurements. Each dataset was reconstructed twice: once using the DOI values predicted by PhysiCal-PET and once using a fixed DOI value of 5~mm for all events. The fixed-DOI reconstruction served as the baseline non-calibrated positioning model. Both datasets were reconstructed with OSEM on a $240\times240\times300$ image matrix with $0.5\times0.5\times0.5~\mathrm{mm}^3$ voxels, using 10 iterations and 5 subsets, without resolution recovery (PSF modeling) so that the measured FWHM reflects the intrinsic resolution gain from the DOI estimates.

The use of PhysiCal-PET DOI estimates improved the reconstructed spatial resolution in all three directions compared with the fixed-DOI reconstruction. The largest improvement was observed in the radial direction, where the mean FWHM decreased from 4.99$\pm0.16$~mm without DOI calibration to 2.90$\pm0.09$~mm with the neural-network-predicted DOI values, corresponding to a 42\% improvement. This substantial reduction is consistent with the expected effect of DOI information in mitigating radial parallax error for off-center source positions.

Improvements were also observed in the axial direction, where the mean
FWHM decreased from $2.03\pm0.34$~mm to $1.60\pm0.10$~mm (21\%). In the
tangential direction the mean FWHM was essentially unchanged
($1.57\pm0.12$~mm vs.\ $1.50\pm0.06$~mm), as expected, since parallax
error primarily affects the radial coordinate for transaxially off-center
sources. These results demonstrate that the DOI estimates obtained from the proposed physics-driven calibration improve image-domain spatial resolution, with the strongest benefit occurring in the radial direction where DOI-related positioning errors have the largest impact.

\subsection{Phantom Reconstruction}
\label{subsec:phantom_reconstruction}

\begin{figure*}[t]
    \centering
    \includegraphics[width=\linewidth]{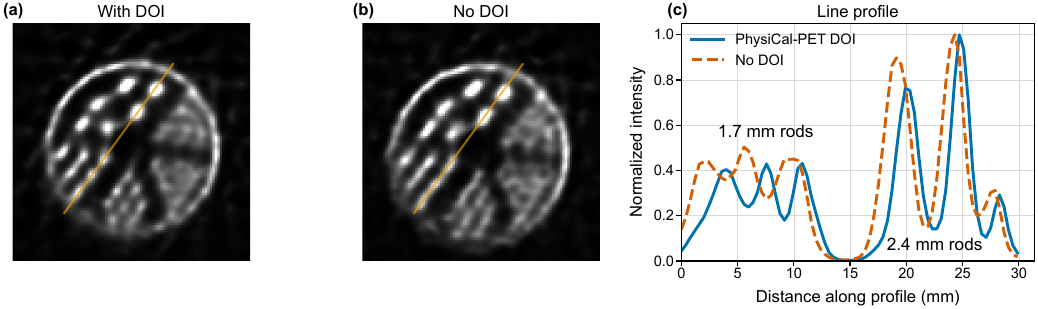}
    \caption{
    Derenzo phantom reconstruction with and without DOI information, averaged across 10 slices. 
    (a) Reconstruction using PhysiCal-PET DOI estimates. 
    (b) Reconstruction without DOI correction, using a fixed interaction depth of 5~mm. 
    (c) Normalized line profiles extracted along the yellow lines shown in (a) and (b). 
    The DOI-based reconstruction shows improved PVR and clearer separation of adjacent rods, particularly in the 1.7~mm rod region, and sharper separation of the 2.4~mm rods. 
    The phantom contains hot rods with diameters of 2.4, 2.0, 1.7, 1.2, 1.0, and 0.7~mm.
    }
    \label{fig:phantom_reconstruction}
\end{figure*}

A Derenzo phantom experiment qualitatively and quantitatively evaluated whether the DOI calibration improves the recovery of small structures. The phantom was placed 36~mm off the radial center and was reconstructed both with the PhysiCal-PET DOI estimates and with a fixed DOI of 5~mm. In contrast to the point-source reconstructions, which used no resolution recovery so that the measured FWHM reflected the intrinsic detector resolution rather than the reconstruction model, both phantom reconstructions applied OSEM on a $300\times300\times375$ image matrix with $0.4\times0.4\times0.4~\mathrm{mm}^3$ voxels, 10 iterations, and 50 subsets, and included resolution recovery (PSF modeling) with a 1.65~mm FWHM kernel. Applying the same kernel to both reconstructions ensures that any difference reflects the DOI estimates rather than the PSF model. Fig.~\ref{fig:phantom_reconstruction} compares the two.

The DOI-based reconstruction shows improved visual sharpness and rod separation compared with the fixed-DOI reconstruction. This improvement is most evident in the smaller rod regions, where the rods appear less blurred and more distinctly separated when the PhysiCal-PET DOI estimates are used. The line profiles extracted along the yellow lines in Fig.~\ref{fig:phantom_reconstruction}(a) and Fig.~\ref{fig:phantom_reconstruction}(b) provide a more direct comparison. In the 1.7~mm rod region, the DOI-based profile shows better separation between adjacent peaks and valleys. Quantitatively, the mean PVR of the 1.7~mm section increased from $1.28 \pm 0.04$ with the fixed DOI to $1.88 \pm 0.33$ with the PhysiCal-PET DOI estimates (a 47\% improvement), with all three valleys meeting the resolution criterion against none for the fixed-DOI reconstruction. The benefit extended to the smaller 1.2~mm rods, where the number of resolved valleys rose from one of five to four of five. The mean PVR of the 1.2~mm section increased more modestly, from $1.35 \pm 0.19$ to $1.48 \pm 0.11$; here the fixed-DOI mean is raised by a single valley with an anomalously deep dip, while its four remaining valleys all fall below the resolution threshold, so the number of resolved valleys summarizes the improvement more faithfully than the section mean. Computing the ratio instead from the mean peak and mean valley amplitude of each section, which is insensitive to such per-valley fluctuations, gave 1.26 versus 1.86 (1.7~mm) and 1.30 versus 1.47 (1.2~mm), consistent with the conclusions above. In the 2.4~mm section both reconstructions recovered the rods, but the DOI-based reconstruction showed sharper peaks and deeper valleys between neighboring rods.

The profile direction is close to the radial direction of the scanner. Therefore, the improved PVR observed in the line profile is consistent with the point-source results, where the largest FWHM improvement was observed radially. These phantom results indicate that the reduction in parallax-induced blurring obtained with PhysiCal-PET DOI calibration can translate into improved recovery of fine structures in reconstructed images.

\section{Discussion}
\label{sec:discussion}

PhysiCal-PET rests on a simple but general principle: if the image is known beforehand, the physical model that maps detector signals to that image can instead be used to calibrate the detector. Conventional PET exploits the annihilation geometry to recover an unknown activity distribution, while the detector response is characterized separately, by normalization, from a well-known homogeneous source~\cite{Badawi1999, Vicente2006}. Here the same idea is extended to intrinsic, event-by-event DOI calibration: by acquiring data from point sources, whose positions can be known essentially exactly, the line-of-response geometry that would otherwise form the image instead constrains the detector property that has not yet been characterized. The present study is a proof of concept that the physics used for image reconstruction can calibrate a fully assembled scanner.

The central practical consequence is that DOI calibration is reduced to a procedure demanding almost no dedicated hardware and minimal human effort: calibrating all 144 detector modules required only an uncollimated point source positioned manually at 11 locations over approximately 55~min of acquisition, with no collimators, motorized stages, benchtop setups, or per-module reference measurements, and the only manual operation being repositioning the source between acquisitions. Because the data are acquired on the assembled scanner, the calibration captures each detector in its operational state, including the final readout electronics and system geometry. The 11 positions used here were chosen to establish feasibility rather than to optimize coverage; a production-grade calibration would likely use more positions so that every crystal is sampled with sufficient statistics and angular diversity, and their optimal number and arrangement remain to be determined.

To our knowledge, the only other recently reported in-system calibration approach that includes the DOI dimension is the 3D in-system calibration method of~\cite{kuhl2025}, which required approximately 8~h to calibrate a single detector, albeit while calibrating all three interaction coordinates rather than DOI alone. In comparison, the proposed method calibrated the DOI response of 144 detectors in under 1~h. A further practical difference is that PhysiCal-PET requires the source position to be localized only once per acquisition, using a simple fixed-DOI reconstruction of the source centroid; this is considerably less demanding than constructing virtual collimation constraints for every source position, and the centroid estimate is statistically robust because it aggregates over millions of events.

The simulation study provides the principal methodological validation. The DOI mean absolute error decreased in parallel with the trimmed source-to-LOR loss and converged to 0.98~mm, showing that the annihilation geometry alone provides sufficient supervision. A stronger form of this validation is provided by the supervised reference model. Training the same architecture directly on the ground-truth DOI yielded a DOI MAE of 0.96~mm, only 0.02~mm below the 0.98~mm obtained by the label-free model. The physics-driven objective reaches essentially the same accuracy as supervised training with perfect labels, which demonstrates that almost no information is sacrificed by replacing explicit DOI labels with the annihilation-geometry constraint. It further implies that the residual error is set by the intrinsic depth sensitivity of the detector signals, rather than by any deficiency of the label-free calibration strategy, and that the geometric loss is a faithful surrogate for the unobservable DOI error. This surrogate property is what enables the experimental application, where no ground-truth DOI labels exist and training can be monitored only through the LOR loss.

In the experimental study, the learned calibration was supported by several independent consistency checks. The predicted DOI values were confined almost entirely to the physical 15~mm crystal depth without any explicit output constraint, with a distribution consistent with the attenuation of 511~keV photons in LYSO. The predicted DOI varied monotonically with the dual-ended signal ratio, as expected from the light-sharing physics of the asymmetric readout; that this relationship differed between blocks even for the same crystal index confirms that a single global DOI--signal mapping would be insufficient and justifies the per-detector and per-crystal conditioning. The calibration also produced a clear image-domain benefit, improving the radial FWHM of reconstructed point sources by approximately 42\% with smaller but consistent gains axially, and the Derenzo phantom showed clearer separation of the rods along a near-radial profile. The peak-to-valley ratio of the 1.7~mm rods rose from $1.28 \pm 0.04$ to $1.88 \pm 0.33$, and the benefit extended to the 1.2~mm rods, four of whose five valleys became resolved with the DOI estimates against one without. Because both phantom reconstructions used the same PSF kernel and the same number of updates, these differences are attributable to the DOI estimates alone; the absolute PVR values, however, depend on the resolution-recovery model and should not be read as a resolution specification of the scanner. The largest improvement was radial, which is consistent with the expected reduction of parallax error, indicating that the learned DOI estimates are geometrically meaningful rather than merely minimizers of the training objective.

The converged experimental trimmed-mean distance-to-LOR (1.14~mm) was larger than the corresponding simulation value (0.26~mm). This gap is consistent with experimental effects that are absent or idealized in simulation, including the finite physical extent of the source, residual uncertainty in the reconstructed source positions, detector and electronic noise, random coincidences, multiple-interaction events, and residual errors in transverse coordinates. It should also be noted that even an ideal calibration would not reach zero, since positron range and photon acollinearity~\cite{MOSES2011S236} impose a physical floor on the achievable source-to-LOR distance.

Several requirements should be stated explicitly: the method needs a correct scanner geometry, and a calibration acquisition with sufficient angular diversity, which motivated the off-axis placement. Its validation is also necessarily indirect, since no ground-truth DOI labels exist, but the agreement among the simulation results, the signal-ratio analysis, and the reconstructed resolution provides converging evidence for the validity of the learned mapping.

Several extensions of this work are envisaged. First, the acquisition time can be reduced further by replacing the sequential single-source measurements with a dedicated calibration phantom containing multiple point sources measured simultaneously. Because the sources are spatially well separated, coincidence events can be assigned unambiguously to their source of origin, and the entire calibration dataset could be acquired in a single measurement of approximately 5~min, reducing the manual effort to a single phantom placement. Second, owing to its speed and simplicity, the same procedure could serve as a routine DOI quality-assurance tool: periodic short acquisitions would allow the DOI response of every detector module to be verified, and drifts caused by temperature variations, SiPM aging, or hardware replacement to be detected and corrected without removing detectors from the system. For verification of an existing calibration, fine-tuning the already-trained network would possibly require substantially fewer counts than training from scratch, although this was not tested in the present work. Third, the PhysiCal-PET framework is not specific to pixelated detectors: the general formulation in Section~\ref{subsec:method_overview} estimates any unknown component of the interaction position, and the approach is expected to generalize to semi-monolithic detectors, where an additional transverse coordinate must be estimated alongside the DOI. Fully monolithic detectors, where all three interaction coordinates are unknown, are also conceivable, although the weaker geometric constraint per event would make the optimization more sensitive and may require additional regularization or acquisition strategies.

The framework is particularly well suited to TB-PET, where the number of detector modules makes conventional DOI calibration most impractical. A larger axial field of view would require a greater number of source positions to cover the imaging volume; however, two properties of TB-PET work in favor of the proposed method. The substantially higher system sensitivity shortens the acquisition time needed to collect sufficient statistics per position, and the abundance of axially oblique coincidences increases the sensitivity of the distance-to-LOR loss to DOI errors, strengthening the very supervision signal on which the calibration relies. The combination of a multi-source calibration phantom with the high sensitivity and angular diversity of TB-PET~\cite{vandenberghe2020state} could therefore enable a fully autonomous DOI calibration of even the largest systems within minutes, which will be investigated in future work.

\section{Conclusion}
\label{sec:conclusion}

This proof-of-concept study demonstrates that the annihilation geometry of PET can replace dedicated DOI calibration hardware. Using only an uncollimated point source placed manually at 11 positions, the DOI response of all 144 detector modules of a preclinical PET scanner was calibrated in 55~min of acquisition, improving the radial point-source FWHM from 4.99 to 2.90~mm (42\%) and raising the peak-to-valley ratio of the 1.7~mm phantom rods from 1.28 to 1.88 while separating the 1.2~mm rods that the uncalibrated reconstruction leaves merged. The framework requires no DOI labels, no collimation, and no mechanical scanning, and its efficiency scales favorably with system sensitivity, providing a path toward autonomous, routine DOI calibration for high-resolution PET systems with a large number of detectors.

\section*{Acknowledgment}
We thank SEDECAL for providing data acquisitions and access to the prototype scanner in order to obtain the measurements required for this study.
\section*{Conflict of interest}
The authors have collaborated with SEDECAL on the design and construction of the prototype scanner reported in this study, and on numerous other PET systems. The measurements, analyses, opinions, and conclusions presented here are solely those of the authors and do not necessarily represent the views or official position of SEDECAL. This article should not be interpreted as an official performance characterization of any SEDECAL product, nor as a guaranteed specification of any commercial SEDECAL scanner. Any performance figures refer exclusively to the prototype configuration described in the text, under the specific acquisition and processing conditions detailed herein, and may differ from those of commercial systems derived from this design.  
\section*{REFERENCES}
\def\refname{References}
\bibliographystyle{IEEEtran}
\bibliography{refs}

\end{document}